# High-performance broadband Faraday rotation spectroscopy of 2D materials and thin magnetic films

*Benjamin Carey[1,4], Nils Kolja Wessling[1,5], Paul Steeger[1], Christoph Klusmann[1], Robert Schneider[1], Mario Fix[2], Robert Schmidt[1], Manfred Albrecht[2], Steffen Michaelis de Vasconcellos[1], Rudolf Bratschitsch[1], and Ashish Arora[1,3]\**

[1]Institute of Physics and Center for Nanotechnology, University of Münster, Wilhelm-Klemm-Str. 10, 48149 Münster, Germany
[2]Institute of Physics, University of Augsburg, Universitätsstr. 1 Nord, 86159 Augsburg, Germany
[3]Department of Physics, Indian Institute of Science Education and Research, Dr. Homi Bhabha Road, 411008 Pune, India
[4]The University of Queensland School of Mathematics and Physics, Saint Lucia, QLD, AUS 4072
[5]Institute of Photonics, Department of Physics, University of Strathclyde, 99 George Street, G1 1RD Glasgow, UK
*Email: ashish.arora@iiserpune.ac.in

We present a Faraday rotation spectroscopy (FRS) technique for measurements on the micron scale. Spectral acquisition speeds of many orders of magnitude faster than state-of-the-art modulation spectroscopy setups are demonstrated. The experimental method is based on charge-coupled-device detection, avoiding speed-limiting components, such as polarization modulators with lock-in amplifiers. At the same time, FRS spectra are obtained with a sensitivity of 20 µrad (0.001°) over a broad spectral range (525 nm – 800 nm), which is on par with state-of-the-art polarization-modulation techniques. The new measurement technique also automatically cancels unwanted Faraday rotation backgrounds. Using the setup, we perform Faraday rotation spectroscopy of excitons in a hBN-encapsulated atomically thin semiconductor $WS_2$ under magnetic fields of up to 1.4 T at room temperature and liquid helium temperature. We determine the A exciton g-factor of $-4.4 \pm 0.3$ at room temperature, and $-4.2 \pm 0.2$ at liquid helium temperature. In addition, we perform FRS and hysteresis loop measurements on a 20 nm thick film of an amorphous magnetic $Tb_{0.2}Fe_{0.8}$ alloy.

**I. Motivation.** When linearly polarized light passes through a material held in a magnetic field applied in the direction of propagation of light, the plane of polarization of the light undergoes Faraday rotation (FR)[1]. It is because the material under a magnetic field offers different complex refractive indices to the circular components ($\sigma^\pm$) of the incident linear polarization[2]. The transmitted light carries rich information on the magnetic-moment-resolved (for instance spin, orbital and valley magnetic moments) band structure of a material[2–7]. Therefore, Faraday-rotation spectroscopy (FRS) is a powerful method in physics, chemistry, and biology. Some notable examples include the magnetic response and domain structures of solids[2,8], optically-detected nuclear magnetic resonances in fluids[9,10], sensitive detection of paramagnetic molecules in gas mixures[11], biochemical and for biomolecular detection[12], spin coherence probing in cold atoms[13,14], investigation of quantum spin fluctuations[15], and laser-frequency stabilization[16]. It is also used to perform Zeeman spectroscopy of many-body quasiparticles in solids, such as neutral and charged excitons[2,17]. As such, it is of fundamental importance to investigate the magnetic response of materials as a function of the photon energy[18]. It is because their characteristic band structures have energy-dependent spin-polarized density of states, and van Hove singularities which possess prominent magnetic responses[18,19]. FRS provides this information with high sensitivity, under magnetic field ($B$) strengths of the order of 1 T or less which are also suitable for device applications[2]. With the recent advancements in the area of two-dimensional (2D) semiconductors and magnets[17,20–26], FRS naturally emerges as a method for studying their magnetic-field-dependent band structures.

A major challenge whilst performing temperature-dependent (typically liquid He temperature up to room temperature) micro-optical spectroscopy on 2D materials is that one has to keep the hBN-encapsulated 2D semiconductors[27–29] spatially stable with submicron precision for the duration of the spectral acquisition[30]. Normally, FRS is performed using modulation-based spectroscopy methods. Such measurements involve a polarization modulator, such as a photoelastic modulator, which periodically varies the state of polarization of the light incident on the sample[31]. The transmitted or reflected monochromatic light is collected using a single-channel detector, such as a photomultiplier tube or a photodiode together with a lock-in amplifier (locked at the fundamental frequency $f$ of the modulator or at $2f$). This technique provides a typical electronic-noise-limited sensitivity of about 35 µrad (0.002°) for measuring the rotation of the state of polarization[31,32]. Since this technique requires scanning the wavelength of the light on the detector, recording an entire spectrum can take several hours. Hence, the spatial stability of the sample position on micron scales in the optical cryogen-flow or closed-cycle cryostats poses a severe challenge in this case over such long measurement times.

In this work, we overcome these challenges and perform high spatial and spectral-precision FRS on 2D semiconductors as well as thin magnetic-alloy films. Given the potential device-based applications of 2D semiconductors, it is imperative to perform Zeeman spectroscopy under device-relevant magnetic field strengths below 1 T. However, so far, existing investigations involve large magnetic fields (> 5 T), because unmodulated methods such as magneto-photoluminescence (PL) or magneto-reflectance are used[17]. Furthermore, since the exciton linewidths in 2D semiconductors such as monolayer $WS_2$, $WSe_2$, $MoS_2$, $MoSe_2$ and $MoTe_2$ are relatively large at room temperature (typically > 20 meV) compared to the Zeeman splittings (typically on the order of 2 meV at $B = 10$ T), almost all measurements are performed at liquid Helium temperatures[17]. At the same time, it is well known from GaAs that the Zeeman splittings of conduction and valence bands can depend strongly on the sample temperature[33,34]. It still needs to be tested if this holds for 2D semiconductors such as monolayer $WS_2$ as well. Here, we perform FR-based Zeeman spectroscopy on a hBN-encapsulated $WS_2$ monolayer at room tempeature and at cryogenic temperature ($T = 10$ K) up to an out-of-plane magnetic field of $B = 1.4$ T and determine the exciton g-factor. As another example, we perform FRS and spectrally-resolved hysteresis-loop measurements on 20 nm thick films of an amorphous $Tb_{0.2}Fe_{0.8}$ alloy to demonstrate capabilities of our method to investigate magnetic materials as well.

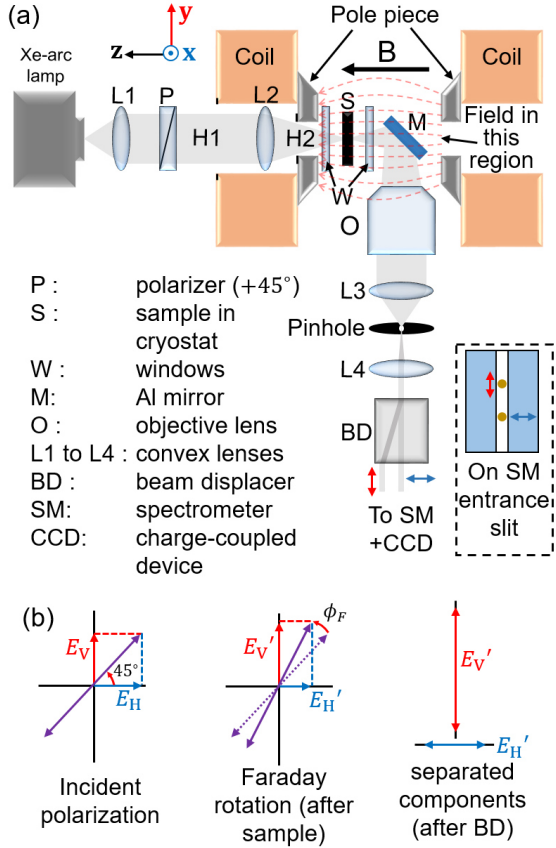

FIG. 1 **Schematic of the Faraday rotation spectroscopy setup.** (a) Schematic (top view) of the Faraday rotation spectroscopy setup depicting the arrangement of the key components. The inset (dashed lines) shows the focused spots corresponding to the two polarization componets split by the beam displacer (BD) at the entrance slit of the spectrometer (SM). (b) Polarization state of the light after passing through the sample and the beam displacer.

**II. Experimental setup.** Our new measurement technique involves a charge-coupled device (CCD)-based signal acquisition, in contrast to conventional polarization-modulation spectroscopy, which employs lock-in amplifiers. Simultaneous wavelength-dispersed data recording of the circular polarizations $\sigma^+$ and $\sigma^-$ by the same CCD chip provides many orders of magnitude higher spectral accummulation speeds (1 – 2 minutes per spectrum) when compared with lock-in-based methods (several hours per spectrum). Additionally, the simultaneous recording of both polarization directions cancels out noise, which is induced by e.g. intensity fluctuations of the light source (similar to an autobalanced detection), yielding shot noise-limited sensitivity. The specific arrangement of the optical elements and our measurement scheme also solve the problem of removing unwanted Faraday rotation backgrounds from the signal. A detailed description of our setup is as follows.

A schematic drawing of the setup is displayed in Fig. 1. Broadband light from a 75 W Xenon arc lamp is collimated and linearly polarized at P = 45° after passing through a sheet polarizer P. It then enters a 30 mm diameter hole H1 from the side of an electromagnet (maximum field 1.4 T, pole gap 47 mm). The light passes through an 80 mm focal length achromatic-doublet lens L2 (400 – 700 nm antireflection coated) installed within the magnet's hole. It is focused on the sample, which is mounted on the extended cold finger of a continuous-flow cryostat, after exiting through a 10 mm diameter hole H2 in the pole piece of the magnet. The sample plane (**x-y** plane) is normal to the direction of the magnetic field (**z**-direction) for performing measurements in Faraday geometry. A 7 mm diameter hole in the the copper cold finger (not shown) lets the light transmit through the sample S. The cold finger is enclosed in a polished non-magnetic stainless steel cold shield and a vacuum shield with two 1 mm thick c-plane (0001) sapphire windows around the sample. After passing through the sample under vacuum, the light undergoes a reflection from an Aluminum mirror M and is steered by 90°. It is then collimated by a long working distance (34 mm) 10x objective lens O (numerical aperture NA = 0.28, focal length 20 mm). A **z**-translation stage attached to the cryostat is used to move the sample to the focal plane of the objective lens. The collimated light is focused on a 20 μm diameter pinhole using a 100 mm focal length achromatic-doublet lens L3 (400 – 1100 nm antireflection coated). The light transmitting through the pinhole is collimated again with another lens L4 (same technical specifications as L3). This arrangement selects light from a spot of 4 μm diameter on the sample. This spatial resolution can be improved further by selecting a smaller pinhole or a larger focal length of L3. The light then passes through a beam displacer (BD), acting as an analyser which spatially splits the beam by 0.6 mm into two orthogonal linearly polarized beams (see Fig. 1b). A 150 mm focal length achromatic doublet lens (not shown) finally focuses the light on a 50 μm wide slit of a 0.3 m focal length spectrometer (Fig. 1a inset). The spectra corresponding to the two polarization directions are imaged onto different rows of a peltier-cooled ($T = 200$ K) charge-coupled device (CCD) mounted at the exit port of the spectrometer and are recorded simultaneously. As described in the supporting information, we measure the signals in two configurations of applied magnetic fields in order to simplify the spectral analysis.

**III. Method**

To explain the functionality of the new Faraday rotation specroscopy setup, a Jones matrix analysis is performed as follows[35]. The Jones' matrices of various components in the setup with magnetic field applied along the +**z** direction ($B^+$) in Fig. 1 following Ref. [32] are:

Incident polarized light: $\hat{P}(p) = E_0 \begin{bmatrix} \cos p \\ \sin p \end{bmatrix}$ (1)

where $E_0$ is the incident electric field.

Sample: $\hat{S}(B^+) = \begin{bmatrix} 1 & -\Phi_F \\ \Phi_F & 1 \end{bmatrix}$ (2)

Mirror: $\hat{M} = \begin{bmatrix} \tan\Psi\, e^{i\Delta} & 0 \\ 0 & -1 \end{bmatrix}$ (3)

Beam displacer: $\widehat{BD}(a) = \begin{bmatrix} \cos^2 a & \cos a \sin a \\ \cos a \sin a & \sin^2 a \end{bmatrix}$, where $a = 0°$ or $90°$. (4)

Here, $\Phi_F = \phi_F + i\eta_F$ is the complex Faraday rotation of the sample, while $\phi_F$ and $\eta_F$ are the Faraday rotation and the Faraday ellipticity[2]. $\tan\Psi$ and $\Delta$ are the ellipsometry parameters of the aluminum mirror[32]. The electric field amplitudes ($E'_H$ and $E'_V$) for the two BD-split polarizations reaching the CCD chip are proportional to

$E'_H \propto \widehat{BD}(0°) \cdot \hat{M} \cdot \hat{S} \cdot \hat{P}(45°)$ (5)

$E'_V \propto \widehat{BD}(90°) \cdot \hat{M} \cdot \hat{S} \cdot \hat{P}(45°)$. (6)

The detected intensities for the two detected polarizations are proportional to $I_H = E'^*_H E'_H$ and $I_V = E'^*_V E'_V$. We define

$\Delta I = I_H - I_V$ and $I = \frac{I_H + I_V}{2}$. (7)

Let us consider that $I_H^0$ and $I_V^0$ are the intensities of the transmitted light for the two polarizations without the sample. Using

$\Delta I^0 = I_H^0 - I_V^0$ and $I^0 = \frac{I_H^0 + I_V^0}{2}$ (8)

we can calculate the following quantity

$\delta(B^+) = \frac{\Delta I}{I} - \frac{\Delta I^0}{I^0} \sim \frac{16\tan^2\Psi\, \phi_F}{(1+\tan^2\Psi)(1+\tan^2\Psi + 2\phi_F(1-\tan^2\Psi))} + \phi_{bg}$ (9)

where, we assume a negligibly small value of $\phi_F$ and $\eta_F$ compared to 1, so that second-order terms in $\phi_F$ and $\eta_F$ are neglected. Here, $\phi_{bg}$ is a background signal which originates from the imperfections in the setup such as i) small strain in the optics and windows leading to linear birefringence, and ii) a slight misalignment in the angles of the polarizer/BD analyser. This background signal can sometimes be quite high compared to the actual Faraday rotation $\phi_F$ [8]. Since, this background signal is independent of the direction of the magnetic field, it can be removed by making two measurements for the opposite directions of the applied magnetic fields i.e. $B^+$ and $B^-$ along +**z** and –**z**, respectively. The Jones' matrix of the sample for $B^-$ is

$\hat{S}(B^-) = \begin{bmatrix} 1 & \Phi_F \\ -\Phi_F & 1 \end{bmatrix}$. (10)

For this case, the Jones' matrix analysis yields

$\delta(B^-) \sim -\frac{16\tan^2\Psi\, \phi_F}{(1+\tan^2\Psi)(1+\tan^2\Psi - 2\phi_F(1-\tan^2\Psi))} + \phi_{bg}$. (11)

We can calculate the Faraday rotation $\phi_F$ from the following quantity

$\delta = \frac{\delta(B^+) - \delta(B^+)}{2} \sim \frac{16\tan^2\Psi}{(1+\tan^2\Psi)^2}\phi_F$. (12)

Hence, the Faraday roration $\phi_F$ can be determined, provided that the $\tan\Psi$ parameter of the mirror is known. For the case of an Aluminum mirror, which we use for our measurements in this work, $\tan\Psi$ is between 0.95 and 0.97 in the wavelength range of $500 - 800$ nm[36]. In this case, Eq. 12 approximately yields

$\delta(\text{Al Mirr}) \sim 3.987\phi_F$ to $3.993\phi_F$ (13)

which is nearly $3.99\phi_F$ within 0.1% over the wavelength range of $500 - 800$ nm.

To summarize the above discussion for obtaining a Faraday rotation spectrum, we have to perform four measurements using the setup in Fig. 1: i) measure $I_H$ and $I_V$ when light passes through the sample (Eq. 7) with $B^+$ applied magnetic field, ii) measure $I_H$ and $I_V$ when light passes through the sample (Eq. 7) with $B^-$ applied field, (iii) measure $I_H^0$ and $I_V^0$ when light passes through air or bare substrate (Eq. 8) with $B^-$ applied field, (iv) measure $I_H^0$ and $I_V^0$ when light passes through air or bare substrate (Eq. 8) with $B^+$ applied field. In each of these steps, light is collected by the CCD for an integration time $t_{int}$, yielding the total measurement time equal to $t = 4t_{int}$. Thereafter, Faraday rotation is calculated using Eq. 12. The Faraday rotation background due to the rotation of light from the optical components inside the magnetic field is eliminated in our method. This is highly advantageous compared to conventional photoelastic modulator and lockin-based FRS or magneto-optic Kerr effect (MOKE) spectroscopy, where the measurement protocol typically does not involve a separate measurement of $I_0$. In those setups, one can have very large FR backgrounds (many degrees) due to thick optics and the substrate directly within the magnetic field[8].

**IV. Quantitative verification of the method.**

To demonstrate the quantitative functionality of our new method based on Eq. 12, we perform the following experiment. In our setup in Fig. 1a, we replace the sample with an achromatic half-wave plate (HWPR) and use it as a calibrated polarization rotator (instead of the sample under magnetic fields). This allows us to precisely set specific rotation values across the entire broad wavelength range, which is key for calibrating the setup. When the HWPR's fast axis is aligned along the direction of incident polarization (at 45° with respect to the **x**-direction), there is no net rotation. However, a rotation of the HWPR by an angle $\theta$ with respect to the incidence polarization results in a rotation by $2\theta$. The HWPR is rotated manually in steps of $\theta = 0.5°$. After each step, the rotation of the incident polarized beam is measured using two methods: the null method (Fig. 2a), and the presented method (Fig. 1). The measurement using the null method is done in steps of 25 nm from $525 - 825$ nm, which is the working range of this broadband HWPR. These results (solid spheres in Fig. 2b) are compared with the results using the presented method (solid lines in Fig. 2b). An excellent agreement between the two methods until $\theta = 3°$ confirms the accuracy of our method. A deviation of rotation from the ideal $2\theta$ value in wavelengths from 525 − 575 nm in Fig. 2b are due to limitations of the HWPR in setting a half-wave retardation accurately in this wavelength region.

*4.1. Influence of ellipsometry parameters of the mirror.* To reveal the effect of the ellipsometry parameter $\tan\Psi$ in Eq. 2, we perform the following measurement. We rotate the HWPR to a certain value (close to 3.2° in this case), which introduces a rotation of 6.2° - 6.4° in the wavelength range from 550 − 700 nm (the working range of this HWPR). The rotation is measured using the method described in Fig. 1, where two types of mirrors M are used: front-polished Aluminum and a polished Si wafer. We plot $\delta/4$ from the two measurements as solid black lines in Fig. 2c and notice that the curve measured when using the Si wafer is lower than that of the Al mirror by a factor lying between $1.12 - 1.16$ (orange line in Fig. 2c). This is explained as follows. We note from Eq. 13 that the expected $\delta/4$ using the Al mirror is approximately $\phi_F$. In the case of the Si wafer, the $\tan\Psi$ parameter lies between 0.69 to 0.7 in the wavelength range of 550 − 700 nm[36]. This yields

$$\delta(\text{Si Mirr}) \sim 3.49\phi_F \text{ to } 3.53\phi_F \quad (14)$$

within the wavelength range. Therefore, $\delta(\text{Al Mirr})/\delta(\text{Si Mirr})$ is expected to be between $\sim 1.13 − 1.15$ which is in excellent agreement with our experimental value. This further justifies the applicability of Eq. 12 to our analysis.

We would like to stress that our measurement protocol also solves a common problem of modulation-FRS or MOKE spectroscopy[32]. In these methods, the presence of a mirror M leads to a mixing of the rotation and ellipticity signals via the ellisometry parameters $\tan\Psi$ and $\Delta$ of the mirror[32]. This requires a non-trivial signal analysis to deduce the actual rotation and ellipticity. In our method, there is no dependence on the $\Delta$ parameter of the mirror, which makes the signal analysis easy compared to the traditional methods[32].

*4.2. RMS noise.* To estimate the root-mean-square (RMS) noise in our measurements, we measure the Faraday rotation of 0.5 mm thick c-cut sapphire in the $525 − 800$ nm wavelength range under a magnetic field of $B = 1.4$ T as a function of integration time. An example of a spectrum obtained in this manner is shown in Fig. 2d for the total measurement time $t = 4t_{int} = 120$ s, where $t$ and $t_{int}$ have been defined in the Method section. RMS noise is calculated in the Faraday rotation spectra as a function of measurement time (spheres in Fig. 2e). The noise follows a $1/\sqrt{t}$ function, elucidating that the measurements are shot-noise limited. The RMS noise in the spectra is found to be better than 0.002° (or 35 µrad) for measurement times $\geq 250$ s. For this short measurement time, this is already comparable to the state-of-the-art modulation spectroscopy techniques[31,32]. For longer accummulation times, the performance gets much better, while still requiring orders of magnitude shorter measurement times compared to the modulation methods. For instance, we acquire FR spectra with 1024 wavelength steps (total CCD pixels in a row) using our method in about 4 minutes. However, such a spectrum with a similar signal-to-noise ratio obtained using modulation spectroscopy would require about 8 − 9 hours (previous experience of authors[32,37,38]), which is more than 2 orders of magnitude longer accummulation time.

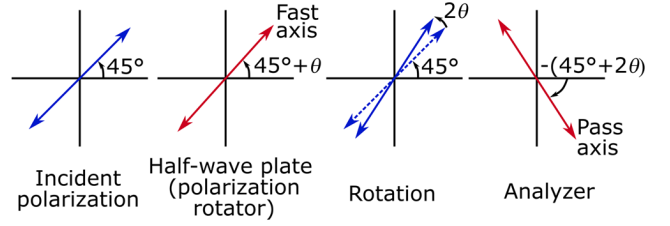
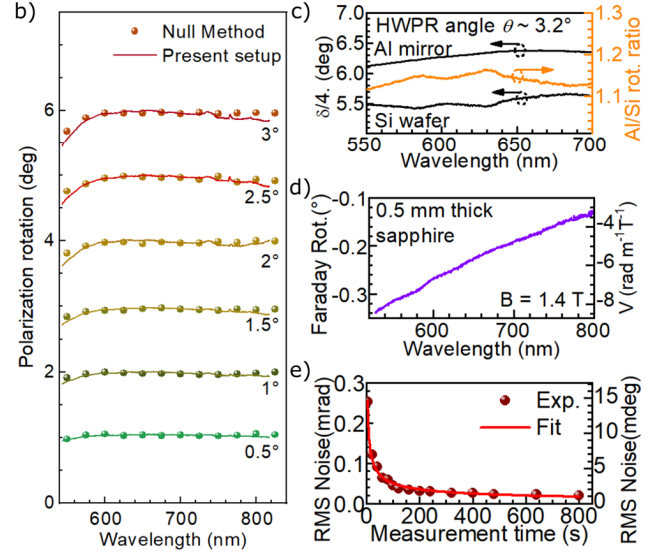

FIG. 2. **Quantitative test of the method**. (a) Optical scheme of the null method used to verify the quantitative working of our setup in Fig. 1. A broadband achromatic half-wave plate (HWPR) is used as a rotator, in place of the sample. (b) Comparison of the optical rotation $2\theta$ introduced by the half-wave plate rotation by $\theta$ using the null method in (a) and the present method. The excellent agreement verifies the applicability of our method. (c) Left axis: $\delta/4$ measured by a setting the HWPR angle $\theta \sim 3.2°$ using an Al mirror, and replacing it with a polished Si wafer. For the Al mirror, $\delta/4$ is nearly equal to the actual value of rotation as explained in the text. Right axis: ratio of rotations measured in the two cases. The factor between 1.12 and 1.16 in the wavelength range of the measurement is in agreement with the theory. The different wavelength ranges in (b) and (c) are due to different HWPRs used in the two cases. (d) An example of a Faraday rotation spectrum of a 0.5 mm thick c-cut double-polished sapphire crystal under a magnetic field of 1.4 T with an integration time of 30s and hence a total measurement time $t$ of 120 s. Similar spectra are measured for other measurement times. The right axis denotes the Verdet constant (V) in units of rad/m/T. (e) Spectra such as in (d) are used to determine root-mean-square noise (spheres) in the Faraday rotation spectra as a function of measurement time. The solid line is the $1/\sqrt{t}$ fit to the data, suggesting that our measurement is limited by shot noise.

**V. Faraday rotation spectroscopy of an hBN-encapsulated WS$_2$ monolayer.** To demonstrate the capabilities of the new Farday rotation setup, we measure the valley Zeeman splitting and valley polarization effects on excitons in an atomically thin transition metal dichalcogenide semiconductor. We prepare an hBN-encapsulated WS$_2$ monolayer on a 0.5 mm thick c-cut sapphire substrate[39]. The room-temperature ($T = 295$ K) and low-temperature ($T = 10$ K) optical transmission spectra of the sample are shown in Figs. 3(a) and 3(e), respectively. The strong response due to the ground-state neutral A exciton $X_A^0$ is identified as a sharp drop in optical transmittance. The full-width at half-maximum (FWHM) linewidth of the exciton resonance is 22 meV and 6.1 meV for 295 K and 10 K, respectively. The corresponding FR spectra are depicted in Figs. 3(b) and 3(f) for magnetic fields from $B = 0.2$ T to 1.4 T (0.2 T $-$ 1.2 T for the low-temperature measurement). We emphasise that such FR spectra would be almost impossible to be measured on these samples using traditional modulation spectroscopy sas in Refs.[37,38], since many hours of sample stability on sub-micron levels is impractical in free-space magneto-optics setups.

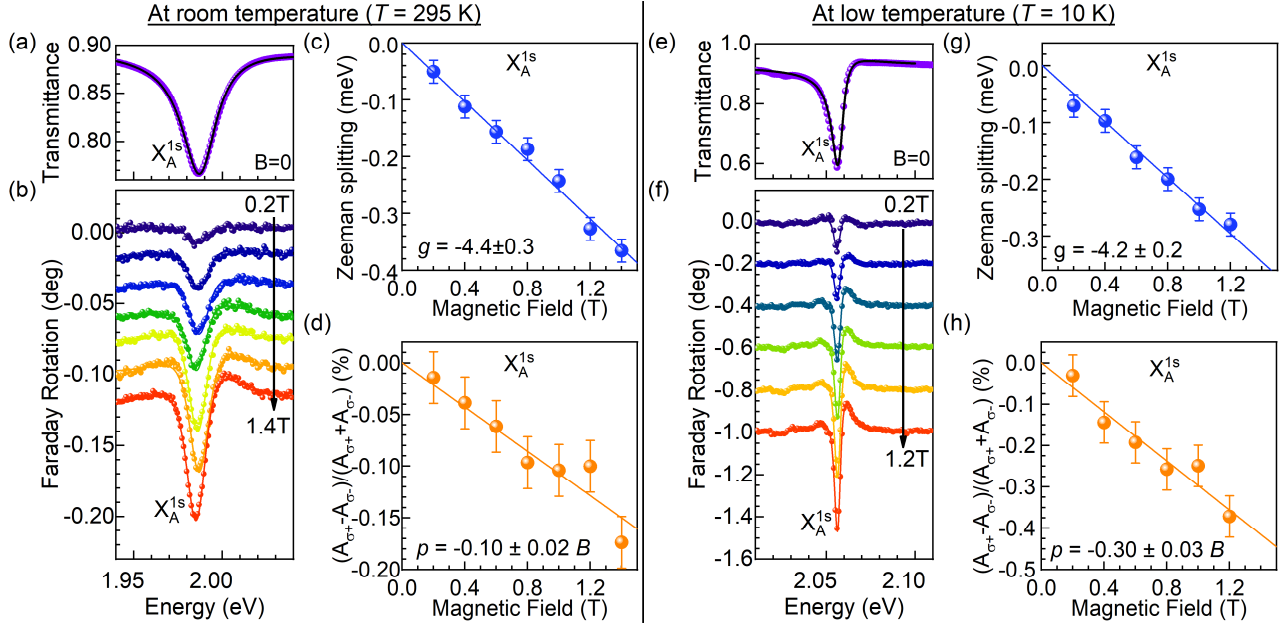

FIG. 3. **Faraday rotation spectroscopy of an hBN-encapsulated WS$_2$ monolayer.** (a) Experimental (purple spheres) and modeled (black solid line) optical transmittance spectrum of an hBN-encapsulated WS$_2$ monolayer for $B = 0$ T at $T = 295$ K. The neutral exciton $X_A^{1s}$ absorbs strongly around 1.988 eV. (b) Experimental (solid spheres) and simulated (solid lines) Faraday rotation spectra of the hBN-encapsulated WS$_2$ monolayer from $B = 0$ T to 1.4 T. (c) Valley Zeeman splitting and (d) magnetic-field-induced valley polarization of $X_A^{1s}$ applying the line shape modeling in (b). Solid lines in (c) and (d) are the fits to the data illustrating the linear change in valley Zeeman splitting and polarization within the field range studied. (e) - (h) are similar to (a) - (d) respectively, but at low temperature of $T = 10$ K. Faint wiggles around 2.03 eV in (e) and (f) originate due to charged excitons[17]. However, they are not analyzed in this work due to their weak signature. The g-factors of $X_A^{1s}$ obtained at $T = 295$ K and $T = 10$ K are $-4.4 \pm 0.3$ and $-4.2 \pm 0.2$, respectively.

The FR spectra are fitted using a transfer-matrix-based analysis technique, similar to the one described in Ref.[37] for MOKE spectroscopy of excitons in GaAs quantum wells: The complex Fresnel transmission coefficients $\tilde{t}_\pm$ of the sample for the circular $\sigma^+$ and $\sigma^-$ polarizations can be written as

$$\tilde{t}_{\sigma\pm} = t_{\sigma\pm} e^{i\theta_{\sigma\pm}}. \tag{15}$$

The Faraday rotation of the light when passing through the sample is related to the phases $\theta_{\sigma\pm}$ as

$$\phi_F = -\frac{1}{2}(\theta_{\sigma+} - \theta_{\sigma-}). \tag{16}$$

To derive $\theta_{\sigma\pm}$, first, the transmittance spectrum is simulated. Assuming that the excitonic contribution to the dielectric function of the material is given by a complex Lorentz oscillator, the dielectric response function of the sample is[37]

$$\epsilon(E) = (n_b + ik_b)^2 + \frac{A}{E_0^2 - E^2 - i\gamma E} \tag{17}$$

where $E_0$, $A$ and $\gamma$ are the exciton transition energy, integrated absorption intensity, and linewidth parameters, respectively. $n_b + ik_b$ is the background complex refractive index of a WS$_2$ monolayer without excitons[40,41]. The fitted transmittance spectra around the A exciton using the transfer-matrix technique are shown as black solid lines in Figs. 3(a) and 3(e). The Fresnel coefficient $\tilde{t}$ is obtained from the fit, and the phase $\theta$ is derived (Eq. 4). An application of a magnetic field leads to a valley Zeeman splitting

$$\Delta E = E_{\sigma+} - E_{\sigma-} = g_X \mu_B B \tag{18}$$

and a valley polarization[17]

$$(A_{\sigma+} - A_{\sigma-})/(A_{\sigma+} + A_{\sigma-}) = \Delta A/2A. \tag{19}$$

Here, $E_{\pm}$ and $A_{\sigma\pm}$ are the exciton transition energies and the oscillator strength parameters, $g_X$ is the effective $g$ factor of the exciton, and $\mu_B = 0.05788$ meV/T is the Bohr's magneton. These effects result in a difference in the phases $\theta_{\sigma\pm}$ and a characteristic Faraday rotation $\phi_F$ spectral line shape (Eq. 16). For instance, a non-zero $\Delta E$ causes a dip in the Faraday rotation spectra at various magnetic fields in Figs. 3b and 3f. The depth of this dip increases with rising magnetic field, since the excitonic Zeeman splitting increases. On the other hand, a non-zero $\Delta A/2A$ results in a shoulder to this dip on the high-energy side. The spectral line shapes are fitted in Figs. 3(b) and 3(f) (solid lines) to determine the magnetic-field-induced exciton Zeeman splitting and polarization. The Zeeman splittings are plotted in Figs. 3(c) and 3(g), while the valley polarizations are plotted in Figs. 3(d) and 3(h). The exciton $g$-factor is found to be $-4.4 \pm 0.3$ and $-4.2 \pm 0.2$ at $T = 295$ K and $10$ K, respectively, after fitting the $B$-dependent Zeeman splitting in Figs. 3(c) and 3(g) with a linear function. The largest source of error in determining these numbers is the fitting error in the line shape modeling of the FR spectra[37,38]. The value of the exciton g-factor is in excellent agreement with previous measurements at liquid He temperatures[17,42–44]. We notice that we do not obtain different g-factors at the two temperatures studied. This is unlike the previously studied bulk GaAs[33,34], and will become a matter of future studies. The exciton valley polarization increases linearly with a slope of $p = -0.1 \pm 0.02\%$ T$^{-1}$ and $-0.3 \pm 0.03\%$ T$^{-1}$ at $T = 295$ K (Fig. 3d) and 10 K (Fig. 3h), respectively. These results highlight the sensitivity of our method, since we are able to measure extremely small Zeeman splittings such as 50 μeV at $B = 0.2$ T in Fig. 3c, with $\pm 20$ μeV accuracy and tiny valley polarizations such as 0.05% at $B = 0.4$ T in Fig. 3d with $\pm 0.02\%$ accuracy. In comparison, other methods reported in the literature such as magnetoreflectance yield limited accuracies of about $\pm 100$ μeV, and $\pm 1\%$ respectively[44–46].

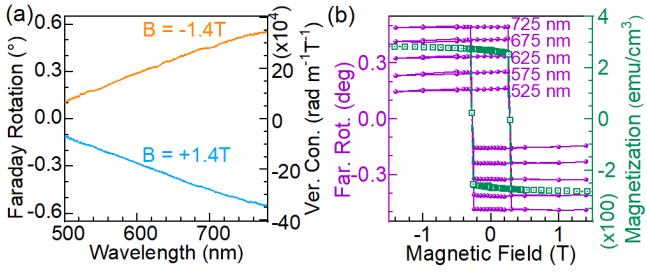

FIG. 4. **Faraday rotation spectroscopy and hysteresis loop of a ferrimagentic Tb$_{0.2}$Fe$_{0.8}$ alloy thin film.** (a) Faraday rotation spectra of a 20 nm thick Tb$_{0.2}$Fe$_{0.8}$ alloy film on a 0.5 mm thick sapphire substrate at a magnetic field of $+1.4$ T and $-1.4$ T. (b) Left axis (purple): Hysteresis loops plotted for 5 different wavelengths to demonstrate the proof of concept. Right axis (green): Hysteresis loop obtained using SQUID - VSM magnetometry showing excellent agreement with the Faraday rotation data.

**VI. Faraday rotation spectroscopy of a Tb$_{0.20}$Fe$_{0.80}$ alloy thin film.** To demonstarate the suitability of our setup in performing state-of-the-art FRS and hysteresis loop measurements on a magnetic film, we measure the Faraday rotation spectrum of a 20 nm thick film of an amorphous ferrimagnetic Tb$_{0.20}$Fe$_{0.80}$ alloy grown on a 500 μm thick c-cut double-side polished sapphire substrate with a 5 nm platinum seed layer. The sample is prepared by magnetron sputtering at room temperature. To avoid oxidation, the layer stack is capped by 5 nm Si$_3$N$_4$. Further details of the sample preparation can be found in Ref.[47]. The sample transmits $7 - 7.5\%$ of the light in the spectral range measured $(500 - 775$ nm$)$ (Fig. 4). At the highest magnetic field available (1.4 T), the magnetization of the film is completely saturated. The magnitude of the Faraday rotation rises almost linearly from 0.1° to 0.55° with increasing wavelength in the investigated spectral range (Fig. 4a). Magnetization hysteresis measurements are performed by sweeping the magnetic field from $+1.4$ T to $-1.4$ T, and back to $+1.0$ T. The measured hysteresis curves at several different wavelengths are shown in Fig. 4b. A sudden reversal of the Faraday rotation at $B = 0.25$ T is observed, leading to a steep hysteresis curve. The data is in excellent agreement with the hysteresis loop measurement, performed by the superconducting quantum interference device - vibrating sample magnetometry (SQUID - VSM) shown in Fig. 4b.

**VII. Conclusion.** In conclusion, we have presented a novel experimental setup for performing Faraday rotation spectroscopy at microscopic spatial resolutions. It is capable of performing measurements with 2 to 3 orders of magnitude enhanced speeds compared to traditional photoelastic-modulator-based setups, while yielding a comparable signal-to-noise ratio. The setup is ideal to perform Zeeman spectroscopy of 2D semiconductors, as well as for hysteresis loop measurements of magnetic thin films. Our work provides a significant advancement in magneto-optics spectoscopy, and opens pathways to optical investigations of spintronic and valleytronic phenomena on micron length scales with high speed and precision.

AUTHOR INFORMATION

Funding Sources

The authors acknowledge the financial support from the German Research Foundation (DFG project nos. AR 1128/1-1 and AR 1128/1-2) and the Alexander von Humboldt foundation.

Notes

The authors declare no competing financial interest.